\documentclass[tightenlines,aps,
nofootinbib,floatfix,nopacs]{revtex4}

\usepackage{psfig,float}
\usepackage{amsmath,amsfonts,graphicx,bm}

\newcommand{\be}{\begin{equation}}
\newcommand{\ee}{\end{equation}}
\newcommand{\ba}{\begin{eqnarray}}
\newcommand{\ea}{\end{eqnarray}}

\begin{document}

\begin{titlepage}

\begin{flushright}
\vbox{
\begin{tabular}{l}
 MADPH-1467\\
 UH511-1090-06\\
 hep-ph/0606237
\end{tabular}
}
\end{flushright}

\thispagestyle{empty}

\title{Fixed target Drell-Yan data and NNLO QCD fits of parton distribution 
functions}

\author{Sergey Alekhin 
}
\affiliation{Institute for High Energy Physics\\ 
Protvino, 142281,  Russia
}

\author{Kirill Melnikov 
}
\affiliation{Department of Physics and Astronomy, University of Hawaii\\
Honolulu, HI, 96822, USA}

\author{Frank Petriello}
\affiliation{Department of Physics, University of Wisconsin,\\
Madison, WI 53706, USA}

\begin{abstract}
We discuss the influence of fixed target Drell-Yan data 
on the extraction of parton distribution functions at next-to-next-to-leading 
order (NNLO) in QCD. When used  in a  parton distribution fit, 
the Drell-Yan (DY) data  constrain   sea quark distributions 
at large values of Bjorken $x$.
We find that not all  available DY data are useful for improving 
the precision of parton distribution functions (PDFs) obtained from a fit to 
the deep inelastic scattering (DIS)  data. 
In particular, some inconsistencies between DIS-based parton 
distribution functions and DY data for large values of 
dilepton rapidity are found. However, by selecting 
a sample of the DY data that is both representative and consistent 
with the DIS data, we are able to perform a combined PDF
fit that significantly improves the precision of non-strange 
quark distributions at large values of $x$.
The NNLO QCD corrections to the DY process 
are crucial for improving the precision.  They reduce the uncertainty 
of the theoretical prediction, making it comparable to the experimental 
uncertainty in DY cross-sections over a broad range of $x$.
\end{abstract}

\maketitle

\end{titlepage}

\section{Introduction}
Parton distribution functions (PDFs) are important 
for the theoretical description of  hard QCD processes at hadron 
colliders. Due  to the factorization of short- and long-distance 
physics,  these functions are universal, and once extracted from one process 
they can be  applied to other hard QCD processes. With the Tevatron Run II under way 
and the LHC upcoming, the need for reliable 
PDFs  is increasing.  Particular aspects that
warrant  careful investigation are PDF uncertainties and their  
influence on theoretical predictions, and the consistent inclusion 
of higher order QCD corrections into PDF fits.

The current standard for perturbative calculations in QCD is 
next-to-leading order (NLO).  The typical accuracy of this approximation 
is  $10-15$ percent.
While this level of precision is
adequate for many physics processes that are studied at the 
Tevatron and will be studied at the LHC, there are 
processes   for which higher accuracy is required. This may happen 
for either calibration processes or important discovery channels, 
such as the production of electroweak gauge 
bosons, the Higgs boson, heavy quarks, and two jets with  large 
transverse momenta.
For such processes, it is desirable to have a theoretical description 
valid through next-to-next-to-leading 
order (NNLO) in perturbative QCD. A significant effort is currently 
under way to develop theoretical tools for  computing parton scattering 
cross-sections with NNLO accuracy.  
To use those calculations for predicting actual hadronic 
cross-sections, parton distribution functions 
 with NNLO accuracy are required as well.

There are currently two distinct approaches to extracting 
PDFs from existing data. The first one is the global fit
that  is practiced 
by the MRST \cite{mrst} and CTEQ \cite{cteq} 
collaborations. The data set in this case includes 
deep inelastic scattering (DIS), Drell-Yan (DY) pair production 
in fixed target and collider experiments, and Tevatron jet cross-sections. 
While such an approach benefits 
from the wealth of data, its drawback is that inconsistent 
data may influence the quality of the fit. In addition,
going beyond the next-to-leading order 
within this framework
is difficult since  very few partonic processes are 
currently known through NNLO in perturbative QCD.

A different  approach to extracting PDFs was suggested in \cite{alekhin}. 
The data set in this case
is restricted to deep inelastic scattering.
Higher order QCD corrections can be included consistently 
within this approach 
since the QCD corrections to DIS
coefficient functions and DGLAP splitting functions 
are known through NNLO \cite{coeff,retey,Moch:2004pa}.
The disadvantage of the DIS-based approach 
is that the DIS data are only sensitive to certain combinations 
of PDFs.  Consequently, not every parton  distribution function can be 
reliably constrained. This leads to  large, approximately 20\%,
errors on sea quark and  gluon distributions
at relatively large values of the Bjorken variable $x$, $x \gtrsim 0.1$.

The determination of sea quark distribution functions can be improved 
if  the approach of 
Ref.~\cite{alekhin} is extended to 
include  precise data on  fixed target 
Drell-Yan processes \cite{e605,e772,e866,towell}.  These data cover the important 
kinematic range 
$Q^2 \sim (20~{\rm GeV})^2$ and $x \gtrsim 0.1$, and are strongly sensitive 
to sea quark distributions in the proton. While such an extension seems 
obvious,  Drell-Yan data was not incorporated into the NNLO  fit
of Ref.~\cite{alekhin} because until recently  
only the NLO calculation of the dilepton rapidity distribution in 
the Drell-Yan process was available \cite{altarelli}. 
Recent NNLO QCD computations \cite{dyrap,zrap} 
of the rapidity distribution  remove this obstacle and permit 
consistent 
inclusion of  the Drell-Yan data in the PDF fit.
The purpose  of the present paper is to perform a combined analysis of the 
DIS and  DY data and to elucidate the impact of the DY data on  parton 
distribution functions.

This paper is organized as follows. In the 
next Section we investigate the consistency of  fixed target DY 
data \cite{e605,e772,e866} and  theoretical predictions 
obtained with the DIS PDFs of Ref.~\cite{alekhin}.
This consistency is the necessary condition 
for combining  the DIS and   DY data; if it is not fulfilled, 
the errors on parton distribution functions  obtained in 
a combined fit are  meaningless. We show that  available DY
data are precise enough so that it is beneficial to include these data in 
a combined DIS/DY fit.
Having established the consistency of the DIS PDFs with  the 
DY data,
we incorporate those data in a combined DIS/DY fit which is 
described in Section III. 
Inclusion of the DY data into the fit 
improves  the precision of sea quark distribution functions 
 for large values of $x$. The quality of the 
DIS/DY fit is similar to the quality of the 
DIS fit of Ref.~\cite{alekhin}. We discuss 
implications of the combined fit for basic QCD and electroweak observables
such as  
the value of the strong coupling 
constant $\alpha_s(M_{\rm Z})$, the Pascos-Wolfenstein ratio and 
the production cross-sections  of  $Z$ and $W$ bosons at the 
Tevatron. Finally, we present our conclusions. 

\section{DIS parton distribution functions  and the DY data}
\label{sec.DIS-DY}

As we discussed in the Introduction, before incorporating 
 fixed target DY data into  the PDF fit based on  the DIS data, we need 
to check if  those data sets are consistent.
To do so, we compute the dilepton rapidity distribution for 
fixed target DY processes using the DIS PDFs \cite{alekhin} and 
compare the results of the calculation to  experimental data
\cite{e605,e772,e866,towell}.
We assume that  dimuon production in the Drell-Yan process
is well 
described by the leading twist factorization and that nuclear 
corrections are unimportant.
There are then two sources of 
 uncertainties in the  
theoretical prediction. First, there is  residual dependence  
on the factorization 
and renormalization scales, a feature common to all 
fixed order calculations in perturbative QCD. 
Second,  parton distribution functions obtained from
a fit to data have systematic uncertainties that influence
the theoretical prediction of the dimuon rapidity distribution.
For the fixed target DY processes that are considered in this paper, 
PDF uncertainties 
are larger than the residual scale uncertainty of the NNLO calculation. 
We are therefore mostly 
concerned with PDFs uncertainties  in what follows.

We choose three sets of fixed target DY  data 
 for our analysis \cite{e605,e772,e866,towell}.
All experiments use an $800~{\rm GeV}$ proton 
beam but employ different targets such as 
hydrogen (E-866), copper (E-605) and deuterium (E-772, E-866). 
The center-of-mass energy of the DY process for 
these three experiments is 
$\sqrt{s} = 38.8~{\rm GeV}$. These experiments therefore cover a broad range of dilepton 
invariant mass $M$ and Bjorken $x$: 
$M \le 20~{\rm GeV}$ and $ x \gtrsim 0.01$. Note that distributions 
in the Feynman variable $x_F$ 
rather than the dilepton rapidity 
 are measured by  E-772 and E-866; however, the  only distribution 
known through NNLO in perturbative QCD is the dilepton rapidity distribution 
\cite{dyrap}.
We relate the $x_F$ distribution and the rapidity distribution 
using   leading order kinematics. This procedure is justified, since for all DY experiments relevant 
for our analysis,  
the average value of the dilepton transverse momentum $p_\perp \sim 1~{\rm GeV}$ is small
compared to the dilepton invariant mass $M \gtrsim 5~{\rm GeV}$. We have checked that 
the use of leading order kinematics  
does not introduce significant bias in the final results.

The sensitivity of parton distribution functions to the DY 
data can be understood from the analytic expression for the DY process 
at leading order 
in perturbative QCD.  The double differential distribution in  dilepton invariant mass $M$ 
and rapidity $Y$ can be written as
\begin{equation}
\frac{{\rm d}^2 \sigma}{{\rm d}M {\rm d}Y}  \sim \sum \limits_{q} 
q_1(x_1) \bar q_2 (x_2) + \bar q_1(x_1) q_2(x_2),
\label{eq1}
\end{equation}
where  $x_1 = M/\sqrt{s}e^{Y}$ and $x_2 = M/\sqrt{s} e^{-Y}$.
Eq.~(\ref{eq1})  
implies that, at leading order, the rapidity distribution is determined 
by either annihilation of a valence quark from the projectile and
a sea antiquark from  the target or  vice versa. 
Valence 
and sea quark distribution functions are determined from the DIS 
data with differing precision.
The precision of valence quark distributions 
is a few percent for all values of  $x$ relevant 
for the  DY and DIS  data that we consider in this paper.  
Sea quark distributions are known from the DIS data with 
a few percent precision only for $x \lesssim  0.1$.  
For larger values of $x$  the error  increases rapidly and exceeds 
20\% \cite{alekhin}. Since  the theoretical 
predictions for ${\rm d}^2\sigma/{\rm d}Y{\rm d}M$ 
are more precise than this error~\cite{altarelli,dyrap},  sea quark 
distributions can be determined 
from Eq.~(\ref{eq1}) with 
an accuracy comparable to    the precision 
of the available   DY fixed target data. 

\noindent
\begin{figure}[htbp]
\vspace{0.0cm}
\centerline{
\psfig{figure=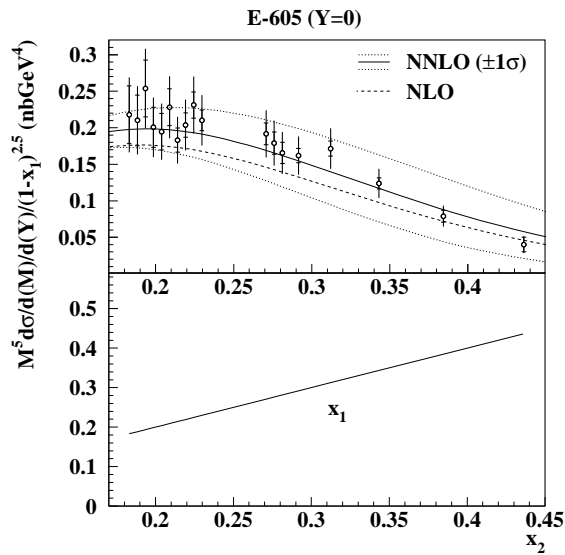,height=8.0cm,width=8cm,angle=0}
}
\caption{
The NLO (dashes) and NNLO (solid) 
dilepton rapidity distributions for proton-copper collisions,
calculated with the DIS PDFs of Ref.~\cite{alekhin}, in comparison
with the E-605 data at zero rapidity.  The NNLO 
$1\sigma$ uncertainty band due to PDF errors is displayed by the dotted curves.  
The relation between $x_1$ and $x_2$ for data points in the upper panel is 
shown in the lower panel.}
\label{e605Y0}
\end{figure}

We begin by comparing theoretical predictions for  
the dilepton double differential distribution in invariant mass and 
rapidity with the E-605 proton-copper scattering
data. The comparison is shown in Fig.~\ref{e605Y0} 
for  the rapidity $Y=0$; note  that 
 {\it different} values of the dilepton invariant 
mass $M$ contribute to this plot. 
In the lower panel of Fig.~\ref{e605Y0},  
values of $x_1$ and $x_2$ are plotted assuming  
leading order kinematics. Theoretical curves are 
computed with the  
NNLO DIS PDFs \cite{alekhin}; 
we choose equal values for the factorization and  renormalization 
scales and  set them equal to the invariant mass 
of the dilepton pair.
The theoretical band reflects  the $1\sigma$ uncertainty of  the DIS PDFs.
It is apparent from Fig.~\ref{e605Y0} 
that for  $x_{1,2} \gtrsim 0.2$, the data are more precise than the 
theoretical prediction 
and the data points are within the theoretical uncertainty 
band. 
The theoretical prediction shown in Fig.~\ref{e605Y0} does not include 
the uncertainty associated with the variation of the renormalization and 
factorization scales. This uncertainty is about ten percent and is much 
smaller than the 30\% PDF error.
It is clear from Fig.~\ref{e605Y0} that the E-605 data are consistent with the 
DIS data, and may therefore be used in the PDF 
fit with  the DIS data.  The precision of the PDFs 
obtained from a combined fit must improve compared to
the situation 
when only the DIS data is fitted.
We note that although Fig.~\ref{e605Y0} refers to a particular 
rapidity value, 
the E-605 data and the theoretical prediction based on the DIS PDFs 
are in agreement for other values of dilepton rapidity as well.

\noindent
\begin{figure}[htbp]
\vspace{0.0cm}
\centerline{
\psfig{figure=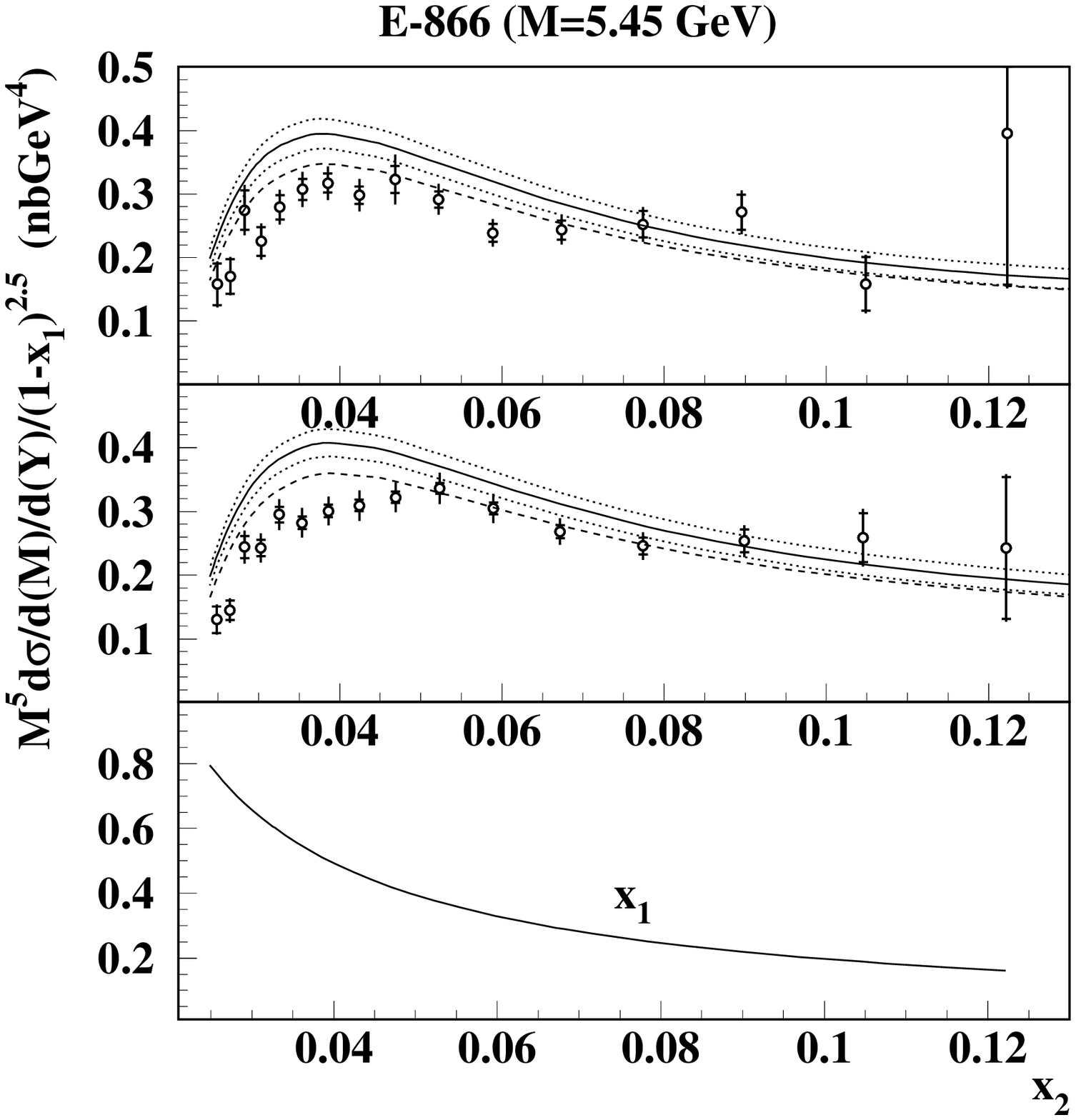,height=8.0cm,width=8cm,angle=0}
\psfig{figure=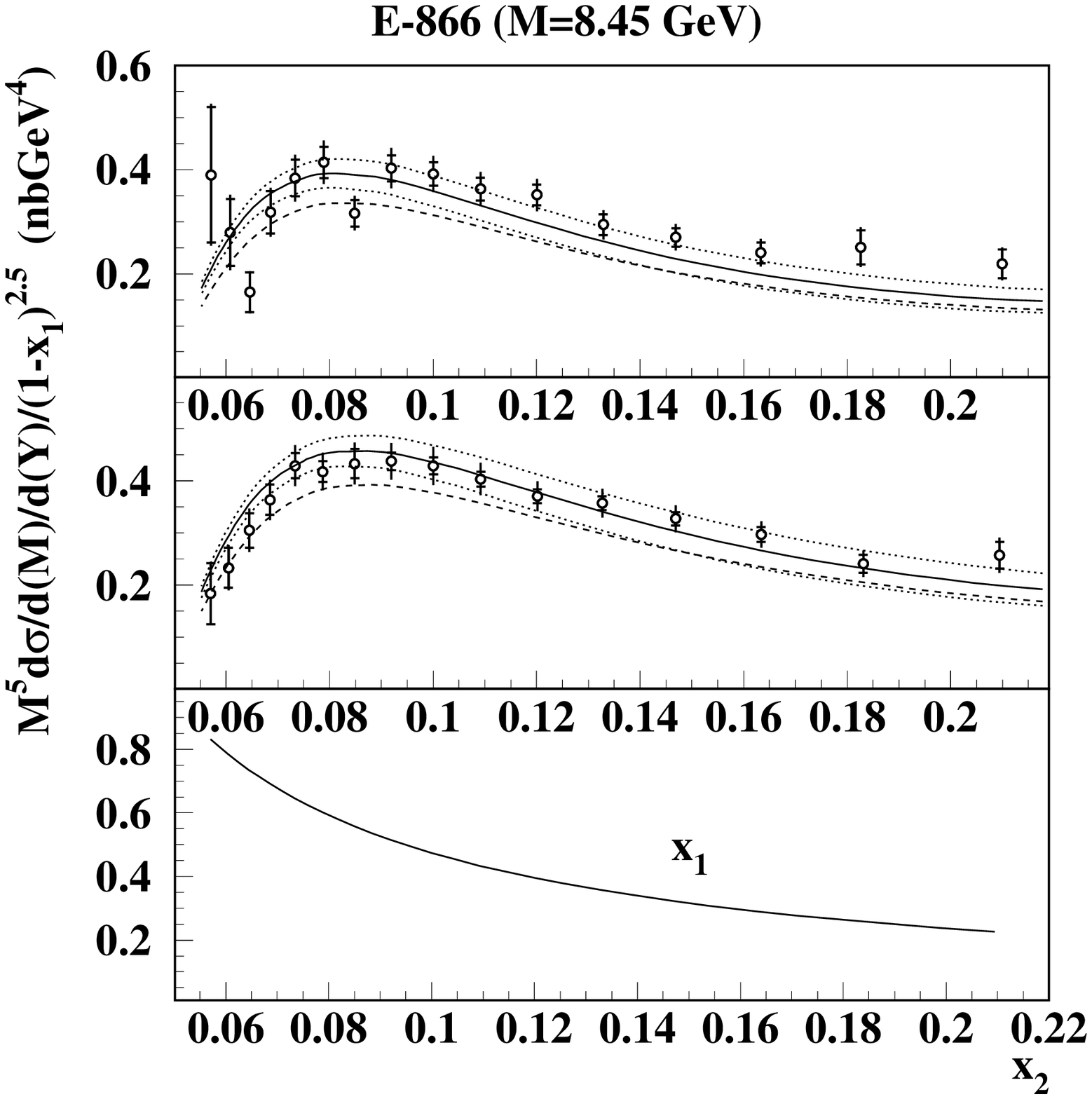,height=8.0cm,width=8cm,angle=0}
}
\caption{The same as in Fig.~\ref{e605Y0} for 
the E-866 proton (upper panels) and deuteron (middle panels)
data for  dilepton masses $M=5.45~{\rm GeV}$ (left panels) 
and $M = 8.45~{\rm GeV}$ (right panels).
}
\label{e866M4}
\end{figure}

A similar analysis can be performed for  the E-866 hydrogen and 
deuterium data; note that the E-866 data covers a broader kinematic 
range than the E-605 data.
In this case, 
we arrive at two different conclusions depending on the invariant 
mass of the dilepton pair produced in the DY process.
We find that for large dilepton invariant masses there is a reasonable 
agreement between  predictions based on  the DIS PDFs and the experimental 
data; this kinematic region is the same as covered by the E-605 data.
However, for small invariant masses  and large rapidities
the E-866 data 
are in  systematic disagreement with theoretical predictions based 
on the DIS PDFs.
The corresponding results are shown in Fig.~\ref{e866M4}.

We now discuss the region of small invariant masses in detail. From 
Fig.~\ref{e866M4} we observe that the experimental data is {\it lower} 
than the theoretical prediction.  
The disagreement occurs in the region $x_1 \gg x_2$ with $x_2 \lesssim 0.1$.
For such values of $x_{1,2}$, $q_{\rm val}(x_1) \sim 
q_{\rm val}(x_2)$ and $q_{\rm sea}(x_1) \ll q_{\rm sea}(x_2)$.  
The second term in Eq.~(\ref{eq1}) is therefore negligible and
the production cross-section   
is mainly  determined by the sea quark distribution 
$\bar q(x_2)$ with $x_2 \lesssim 0.1$. 
However, for such values of $x_2$ the 
precision of  sea 
quark distribution functions obtained from the DIS data is 
close to a few percent \cite{alekhin}. We therefore conclude that  
for this kinematic range, the available DY data 
can not improve the precision of the DIS PDFs. 
Instead, the theoretical prediction for the dilepton rapidity distribution 
obtained with  the DIS PDFs is a non-trivial check of the consistency of 
the data. It follows from  Fig.~\ref{e866M4} that this 
consistency check fails since the experimental data are systematically below theoretical 
predictions.  We note that the NLO theoretical prediction is in better agreement 
with the data.  While this is clearly accidental, it may result in misleading 
conclusions about the compatibility of different data sets.
Forcing PDFs  to fit {\it both} data sets is a bad solution\footnote{Note that 
this is exactly what happens in available global fits.}; the 
PDFs obtained in that case 
result in  rapidity distribution curves that pass 
{\it between}  the DIS-based prediction and the E-866 data, 
the fit quality  deteriorates 
and no reduction of the PDF uncertainty is achieved. 
We conclude that there is a contradiction between the 
DIS data  and the small dimuon mass data obtained by the E-866 collaboration. 
In the region where the disagreement occurs, the PDFs 
are already known precisely from the DIS data. Hence, for such values of 
dilepton invariant 
mass the DY 
data  does not improve the 
precision of  sea quark PDFs.

\begin{figure}[htbp]
\vspace{0.0cm}
\centerline{
\psfig{figure=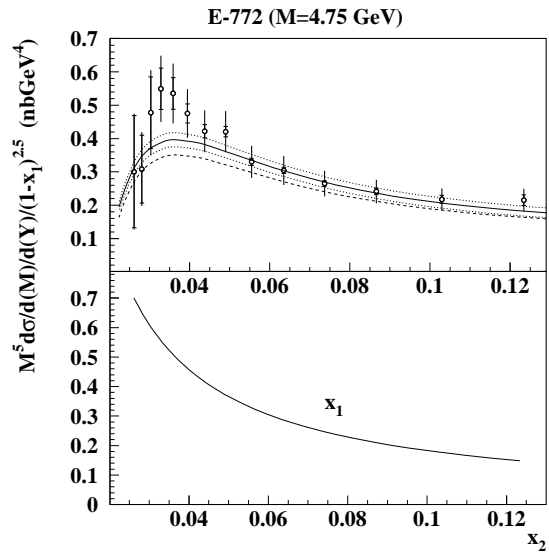,height=8.0cm,width=8cm,angle=0}}
\caption{Te same as in Fig.~\ref{e605Y0} for the
E-772 deuterium data and for the dimuon invariant mass $M=4.75~{\rm GeV}$.
}
\label{e772M4}
\end{figure}

The disagreement between the DIS-based prediction and the 
E-866 data for small invariant masses occurs at large rapidities.
This kinematic region is known to be problematic for 
existing fixed target DY experiments. 
In particular, there is a disagreement between the E-772 and E-866 deuterium
data, with the E-772 data points being systematically higher.  
In principle, this is exactly 
what is needed to match the DIS-based prediction and the DY data, as can be seen from Fig.~\ref{e866M4}. However, as shown in 
Fig.~\ref{e772M4}, 
the E-772 data points are  somewhat too high on average. 

\noindent

We suspect that problems with  the large rapidity region
originate  from underestimated systematic uncertainties.
If this is the case, the ratio of 
cross-sections for hydrogen and deuterium targets measured 
by  the E-866 collaboration \cite{towell} is useful since many  
systematic uncertainties cancel in the ratio. 
We note that the    
theoretical prediction for the ratio is also more precise; for example, the dependence on the 
factorization and  renormalization scales, a ten percent effect in the 
theoretical predictions for individual cross-sections, 
disappears in the ratio.

\begin{figure}[htbp]
\vspace{0.0cm}
\centerline{
\psfig{figure=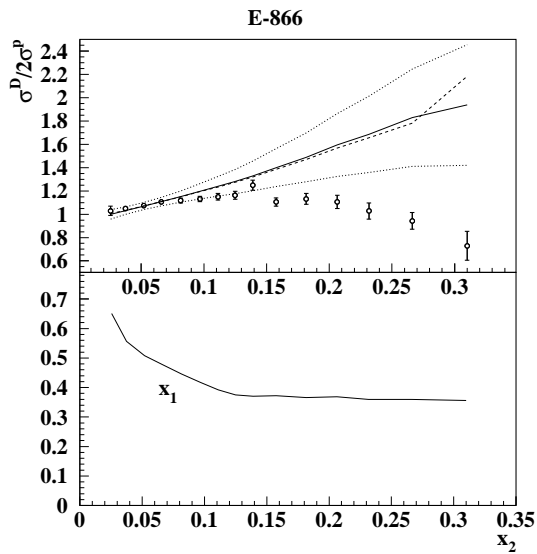,height=8.0cm,width=8cm,angle=0}
}
\caption{The same as in Fig.~\ref{e605Y0} for the deuteron to proton 
cross-section ratio
measured by the E-886 experiment. Larger values of $x_2$ 
correspond to larger dimuon invariant masses.
}

\label{e866pd}
\end{figure}

The E-866 results for the ratio of deuteron to proton 
cross-sections
and the theoretical prediction based on the DIS PDFs 
are compared  in Fig.~\ref{e866pd}. 
In this case, there is an 
agreement between theory and data for small invariant 
masses, whereas for larger invariant masses and larger values of Bjorken $x$, 
the shape of the DY data differs from the DIS prediction. However, 
this region is not really problematic for the consistency of the 
DIS and DY data  since it is strongly sensitive to sea 
quark PDFs for $x \gtrsim 0.1$, where 
sea quark PDFs obtained from the DIS fit 
suffer from large  uncertainties \cite{alekhin}. Given the large PDF errors 
in the region  of $x$ where the disagreement occurs, 
we conclude that the E-866 data on the ratio of  
deuteron to proton cross-sections  can be used together with 
the DIS data without sacrificing the quality of the fit.

Having compared  theoretical predictions  based on 
the DIS PDFs with  the data on DY processes, we briefly discuss   changes that
can be expected once the DY data is included in the fit.
As an illustration, 
consider the E-866 data for the dimuon invariant mass $M=8.45~{\rm GeV}$, 
shown in Fig.~\ref{e866M4}. For larger values of $x_2$, we observe that for both 
proton and deuteron targets the experimental data points 
are somewhat higher than the theory prediction. To make  theory agree 
with experiment, we require that sea quark distributions for $x \gtrsim 0.1$ 
 {\it increase}. Moreover, since  the disagreement between theory and 
experiment is stronger for the proton data, the $\bar u$ distribution function
should receive a larger increase than the $\bar d$ distribution. This observation is consistent 
with the results for the ratio of  deuteron to proton cross-sections in 
Fig.~\ref{e866pd}. The ratio of the two cross-sections can be approximated by 
\begin{equation}
\frac{{\rm d}\sigma(pd)}{2{\rm d}\sigma(pp)}|_{x_1 \gg x_2}
 =\frac{1}{2} \left (
1 + \frac{\bar d(x_2)}{\bar u(x_2)}
\right ).
\label{eq2}
\end{equation}
It follows that since the  
ratio of  {\it computed} cross-sections is {\it higher} 
than the experimental result, the ratio $\bar d/\bar u$ should 
decrease.  This  is consistent with the information from the 
absolute measurement 
of proton-proton and proton-deuteron
cross-sections.
It is interesting to 
note that the $\bar d$ distribution function 
 almost coincides for DIS PDFs \cite{alekhin} 
and MRST PDFs \cite{mrst}, whereas   the $\bar u$ distribution function 
from the DIS fit is smaller 
than the one obtained by MRST. This is not accidental, since MRST includes 
the E-866 data in their fit.  While the preceding discussion indicates 
how sea quark distributions are influenced by the DY data, it  
 is less obvious that gluon PDFs at large values of  
$x$ may also be affected. To see that this may happen, 
recall that the contribution of the $qg$ partonic subprocess  
to the dimuon production cross-section 
is relatively large, approximately 15\% of the total, 
and  {\it negative}. Decreasing
the gluon content of the proton may
therefore increase the rapidity distribution.  A 
similar effect can be achieved by increasing sea quark distributions.  
Since both gluon and sea PDFs at large $x$ are poorly constrained by the 
DIS fit, the impact of the DY data on each of these distributions can not 
be disentangled using qualitative considerations.

\noindent

Following the discussion 
in this Section, we include 
the E-605 data and the  E-866 data on the ratio of  deuteron to proton
cross-sections in the combined DIS/DY 
fit. These two data sets improve the precision 
of sea quark distributions obtained from the DIS fit in two different ways.
The E-605 data improves the precision of 
sea quark distributions for $x \gtrsim 0.2$ in a ``flavor-blind'' fashion, 
whereas $\bar u - \bar d$ for $x \gtrsim 0.1$
is constrained by the E-866 data. Note that 
even if the E-866 and E-772 measurements of  
the absolute proton and 
deuteron cross-sections were consistent with the DIS data, 
they could not have  added much new information
compared to the DY data which we include in the fit. 
This is because at small rapidities
the E-605 data is as good as the E-866 and E-772 data, while at large 
rapidities the  PDFs are already constrained by the DIS data.
We conclude that our selection of the DY data is sufficiently 
representative  
and can be combined with the DIS 
data to determine parton distribution functions 
with high  precision.
We describe the results of the  combined fit in the  next Section.

\section{A fit to the combined DIS and DY data}
\subsection{Theoretical input}

In this Section we fit PDFs to both the DIS and DY data.  
We begin with a brief description of the salient features of the approach 
in Ref.~\cite{alekhin}. We use the following  parameterization of 
parton distribution functions 
at $Q^2_0=9~{\rm GeV}^2$:
\begin{eqnarray}
&& xq_{\rm V}(x,Q_0)=\frac{ 2 \delta_{q{\rm u}}+\delta_{q{\rm d}}}
{N^{\rm V}_q}x^{a_{q}}(1-x)^{b_q}x^{P_{q,\rm V}(x) },
~~~P_{q,{\rm V}} = \gamma_{1,q} x + \gamma_{2,q} x^2,
~~q = u,d;
\label{eqn:pdf1}
\\
&& 
xq_{\rm S}(x,Q_0)=A_q x^{a_{q{\rm s}}}(1-x)^{b_{q{\rm s}}}
x^{P_{q,\rm s}(x) },~~~~~~~~~~~~~~P_{q,{\rm s}} = \gamma_{1,{q{\rm s}}} x,~~~~~~~~~~~~~~~~~~~~~q = u,d,s;
\label{eqn:pdf3}
\\
&& xG(x,Q_0)=A_{\rm G}x^{a_{\rm G}}(1-x)^{b_{\rm G}}
x^{P_{\rm G}(x) },~~~~~~~~~~~~~~~~P_{\rm G} = \gamma_{1,{\rm G}} x.
\label{eqn:pdf5}
\end{eqnarray}
Valence quark distributions are displayed in Eq.~(\ref{eqn:pdf1}), 
sea quarks are shown in Eq.~(\ref{eqn:pdf3}),  and  
gluons are shown in Eq.~(\ref{eqn:pdf5}).
To obtain PDFs at arbitrary $Q^2$, we  employ 
the DGLAP evolution equation with the NNLO Altarelli-Parisi splitting 
kernels computed recently  \cite{Moch:2004pa}.
The PDF parameterization in Eqs.~(\ref{eqn:pdf1}-\ref{eqn:pdf5})
differs from the one in Ref.~\cite{alekhin}.  
It allows more flexibility, which is important since more data 
are included in the fit.
Note that some  parameters in Eqs.~(\ref{eqn:pdf1}-\ref{eqn:pdf5})
are inter-dependent. For valence quarks, $N_V^{q}$ is calculated 
from the requirement that the total numbers of valence $u$ and $d$ quarks 
are two and one, respectively. Also, the normalization of the gluon 
distribution, $A_{\rm G}$, is related to the 
other parameters through the  momentum conservation constraint. Since 
the strange quark distribution is not well constrained by the data 
used in the fit, we fix it using  the
CCFR data on dimuon production in  neutrino-nucleon collisions
\cite{Bazarko:1994tt}. This leads to 
$A_s = 0.08$, $b_{s{\rm s}} = 7$, and $\gamma_{1,s{\rm s}} = 0$.
We also set  $a_{us}=a_{ds}=a_{ss}$ which is a natural choice since the  
existing
DIS/DY data is not useful for detecting  non-universality of sea PDFs
at small $x$.
The contribution of heavy quarks to  DIS structure functions 
is accounted for within the massive factorization scheme 
using the one-loop  computations  of Ref.~\cite{Laenen:1992xs}.
For the fixed-target DY data employed  in the fit, 
heavy quark contributions  are  unimportant.

\noindent
\begin{figure}[htbp]
\vspace{0.0cm}
\centerline{
\psfig{figure=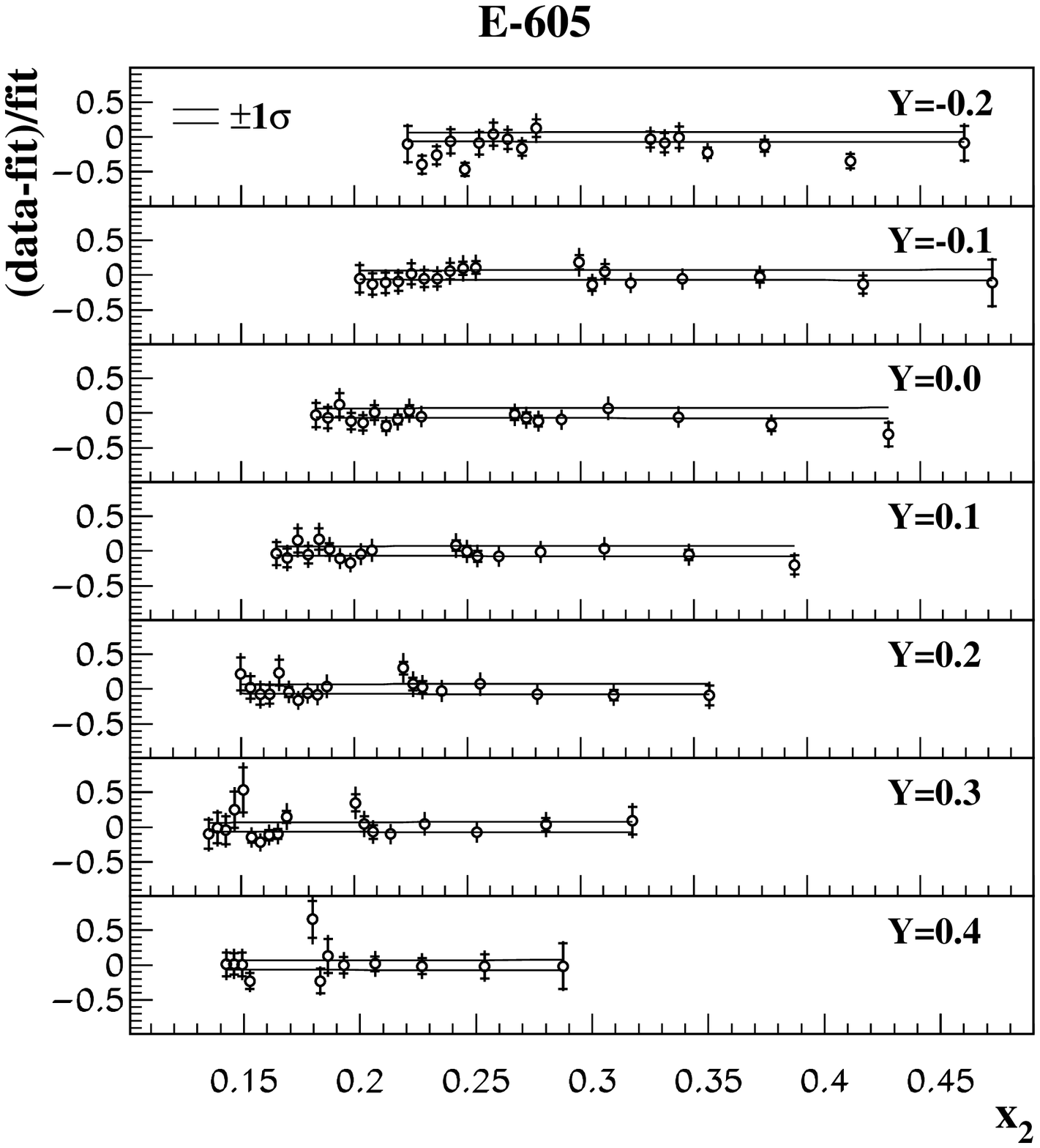,height=8.0cm,width=8cm,angle=0}
\psfig{figure=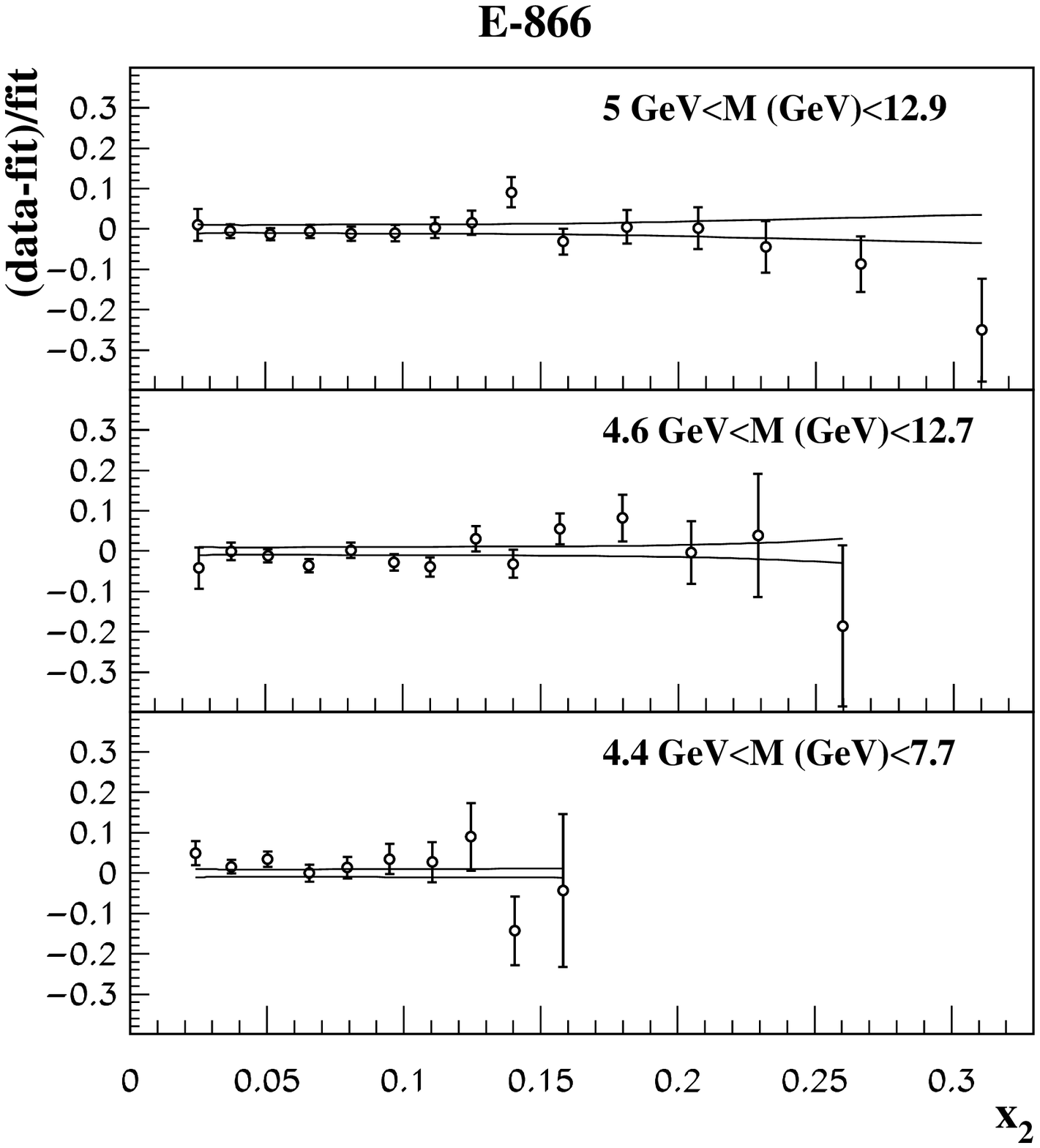,height=8.0cm,width=8cm,angle=0}
}
\caption{Data points used in the DIS/DY fit vs. 
predictions based on fitted PDFs. 
The bands reflect  the   $1\sigma$ uncertainty of fitted PDFs.
}
\label{e605_fit}
\end{figure}

The DIS
deuteron data are corrected for  nuclear effects that
include Fermi motion, shadowing and  nucleon off-shellness. 
Since  the deuteron nuclear correction increases with $x$,
the cut $x < 0.75$ was applied to the DIS deuteron target data in 
Ref.~\cite{alekhin}.  Because uncertainties  in 
nuclear effects at large $x$ are now better understood~\cite{Kulagin:2004ie},  
we do not apply a similar cut in the current analysis 
and include the DIS data points
up to $x=0.9$, the largest value of $x$ available in the existing 
DIS data. 
The DY data are not corrected for  nuclear effects since
these data  points 
are concentrated at $x \lesssim 0.3$, where nuclear corrections 
are small~\cite{Alde:1990im}.

Our treatment of  power corrections to  logarithmic evolution of the 
DIS structure functions follows Refs.~\cite{alekhin,Alekhin:2003qq}. 
We suppress the sensitivity of the structure functions to  power-like terms
by removing  the DIS data with $Q^2<2.5~{\rm GeV}^2$ 
and hadronic invariant mass $W<1.8~{\rm GeV}$.
For the remaining data, target mass corrections important at large $x$ are  
applied using the Georgi-Politzer scheme~\cite{Georgi:1976ve}.
Applying just the target mass corrections is insufficient.  We must also 
add  twist-4 terms to the DIS structure functions. 
These terms  are parameterized by cubic spline polynomials of $x$ 
whose  coefficients are fitted to data.
Note that twist-4 contributions produce only $\sim 10\%$ 
corrections to DIS PDFs even for $Q^2 \sim 2.5~{\rm GeV}^2$ and 
become unimportant for $Q^2 \sim 20~{\rm GeV}^2$. By analogy, 
since the DY data employed in our analysis correspond to
$Q^2 \ge 25~{\rm GeV}^2$, we do not consider power corrections 
to this part of the data sample.

\begin{table}
\caption{The number of data points (NDP) and  values of $\chi^2/{\rm NDP}$
for each experiment used in the fit.}
\begin{center}
\begin{tabular}{|c|c|c||c|c|c|}
\hline
Experiment  &NDP&$\chi^2/{\rm NDP}$ & 
Experiment  &NDP&$\chi^2/{\rm NDP}$ \\  \hline
SLAC-E-49A  &118&0.56 & BCDMS       &605&1.10\\  \hline 
SLAC-E-49B  &299&1.18 & NMC         &490&1.26\\  \hline
SLAC-E-87   &218&0.94 & H1(96-97)   &135&1.13 \\ \hline 
SLAC-E-89A  &148&1.42 & ZEUS(96-97) &161&1.28\\  \hline
SLAC-E-89B  &162&0.80 & FNAL-E-605        &119&1.49 \\ \hline
SLAC-E-139  &26 &1.03 & FNAL-E-866        &39 &1.13\\  \hline 
SLAC-E-140  &17 &0.47  & Total                & 2537 &  1.13  \\ \hline
\end{tabular}
\end{center}
\label{tab.chi}
\end{table}

\subsection{Results of the fit}

The PDF parameters in Eqs.~(\ref{eqn:pdf1}-\ref{eqn:pdf5}) and the 
coefficients of the twist-4 corrections  to  the DIS structure functions 
 are obtained from the fit to the 
DIS data for proton and deuteron targets~\cite{Whitlow:1992uw} 
and  the DY data of Refs.~\cite{e605,Towell:2001nh}. 
To check that our PDF parameterization is sufficiently flexible,  
we modified the polynomials $P_{q,G}(x)$ in 
Eqs.~(\ref{eqn:pdf1}-\ref{eqn:pdf5}) by adding terms of the type 
$\gamma_n x^n$, $n =2,3$ for the sea, gluon and valence  distributions.  
We found
that such modifications  do not improve the description of 
the data. The overall quality of the fit is good; 
for its final version  the value 
$\chi^2/{\rm NDP}=2862/2537=1.13$ is obtained.

\noindent
\begin{figure}[htbp]
\vspace{0.0cm}
\centerline{\hbox{
\psfig{figure=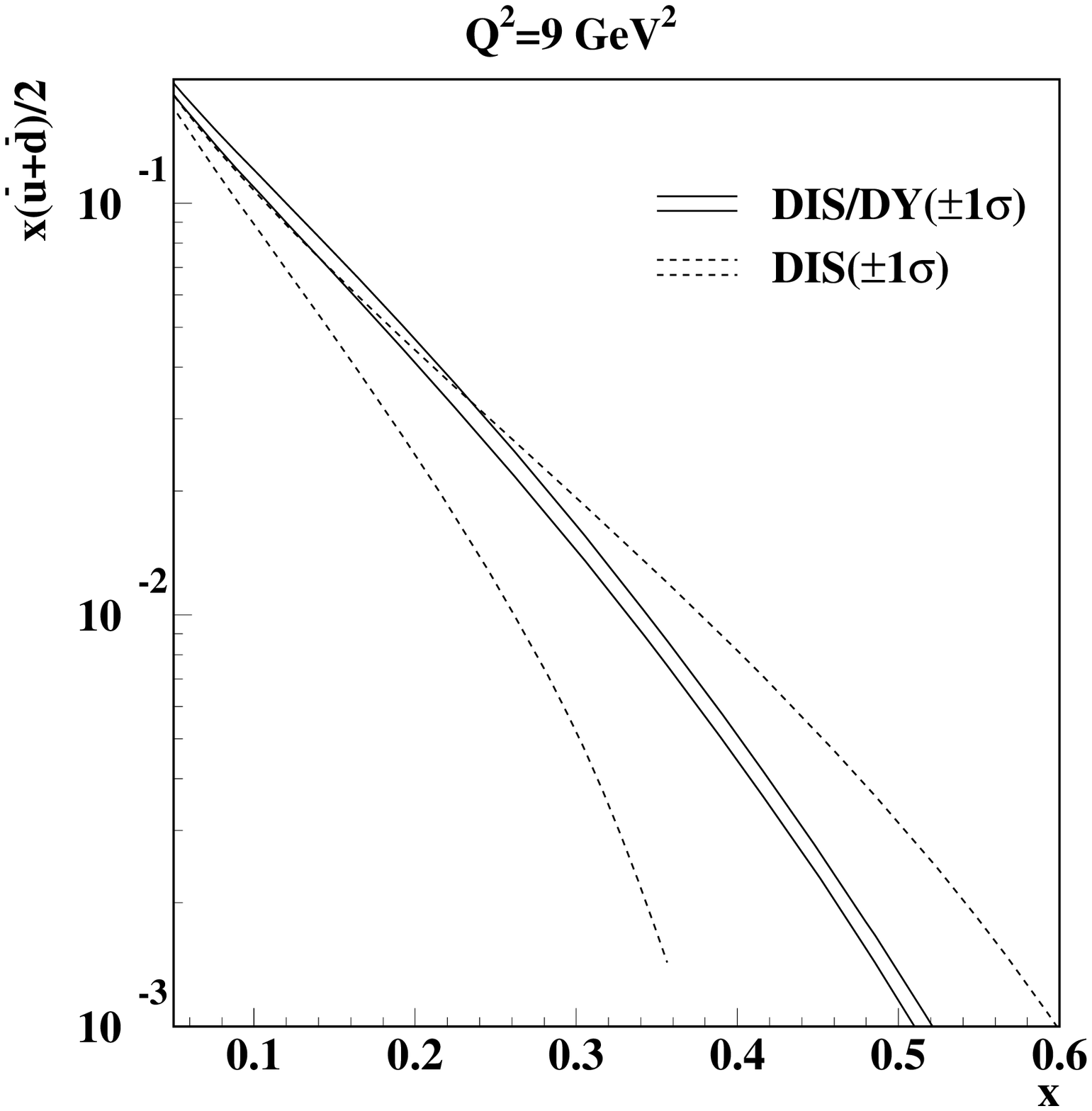,height=6.0cm,width=8cm,angle=0}
\psfig{figure=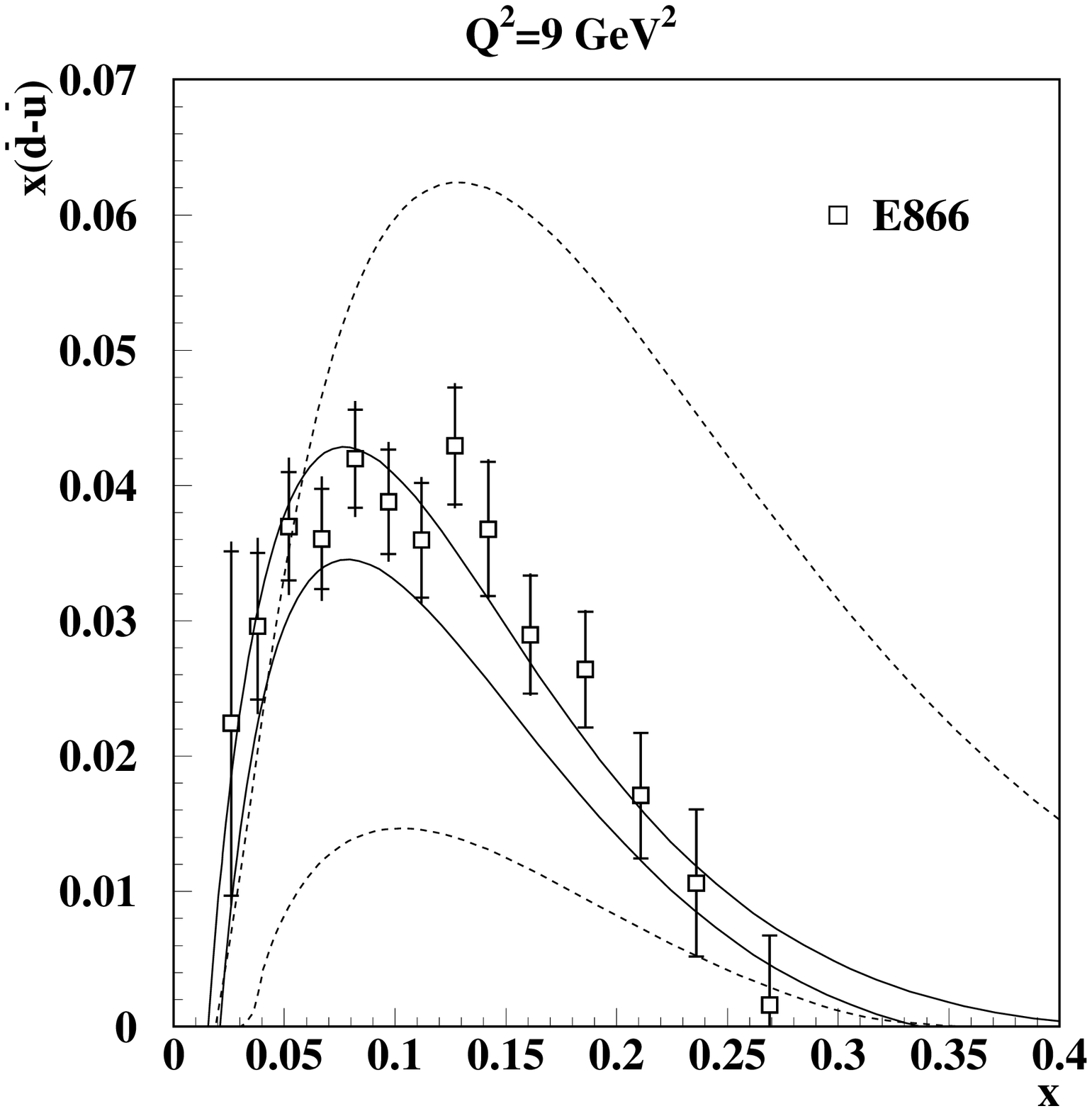,height=6.0cm,width=8cm,angle=0}
}}
\caption{The 1$\sigma$ bands for
isospin-symmetric and anti-symmetric sea quark distributions
from the DIS/DY (solid) and  DIS (dashes) fits.
}
\label{qs}
\end{figure}

To demonstrate the quality of the fit in more detail, we show  
values of $\chi^2/{\rm NDP}$ for   separate experiments 
in Table~\ref{tab.chi}. It is clear from the Table that the description of the 
data is acceptable.
In Fig.~\ref{e605_fit} results for the pulls of the DY  
data used in the fit
are displayed.  They do not demonstrate any systematic trend. 
The description of the  E-605 data has randomly distributed deviations
that can be attributed to 
fluctuations beyond quoted experimental errors. 
We can model the possibility of some experimental errors being underestimated 
by re-scaling the errors for 
experiments with $\chi^2/{\rm NDP} > 1$.
We find that these scale factors do not exceed 
 1.2 and the impact of  the 
re-scaling on the PDF errors is within   20\%.

\begin{table}
\caption{Parameters of parton distribution functions derived from the 
NNLO QCD fit to the DIS and DY data.
The errors on fit parameters are obtained by propagating  the 
statistical and systematic errors in the data. As described in the text, 
$a_{u \rm s}$ and $a_{d \rm s}$ are identical by construction.}
\begin{center}
\begin{tabular}{|c|c|c|c|c|c|}   
\hline
 &  $u_v$ & $d_v$ & $u_s$ & $d_s$ & $g$ \\ \hline
$a$ & $0.670 \pm 0.035$  & $0.61\pm 0.12$ & 
$-0.2182\pm 0.0044$&  $-0.2182\pm 0.0044$ 
& $-0.198\pm0.015$ \\ \hline
$b$ & $3.639 \pm 0.077$ & $5.21 \pm 0.42 $ & $6.14 \pm 0.25 $ & 
$8.24 \pm 0.40 $ & $5.41 \pm 0.13 $ \\ \hline
$\gamma_1 $ & $-0.41 \pm 0.27$ & $0.18 \pm 0.27 $ & 
$1.04 \pm 0.32 $ & $-1.97 \pm 0.48 $ & $2.09 \pm 0.94 $ \\ \hline
$\gamma_2 $ & $-0.91 \pm 0.18$ & $-4.19 \pm 0.18 $ & 
 &  &  \\ \hline
$A $ &  &   & $0.1488 \pm 0.0060 $ 
 & $0.1220 \pm 0.0063 $ &  \\ \hline
\end{tabular}
\end{center}
\normalsize
\label{tab:pdfpars}
\end{table}

\noindent
\begin{figure}[htbp]
\vspace{0.0cm}
\centerline{
\psfig{figure=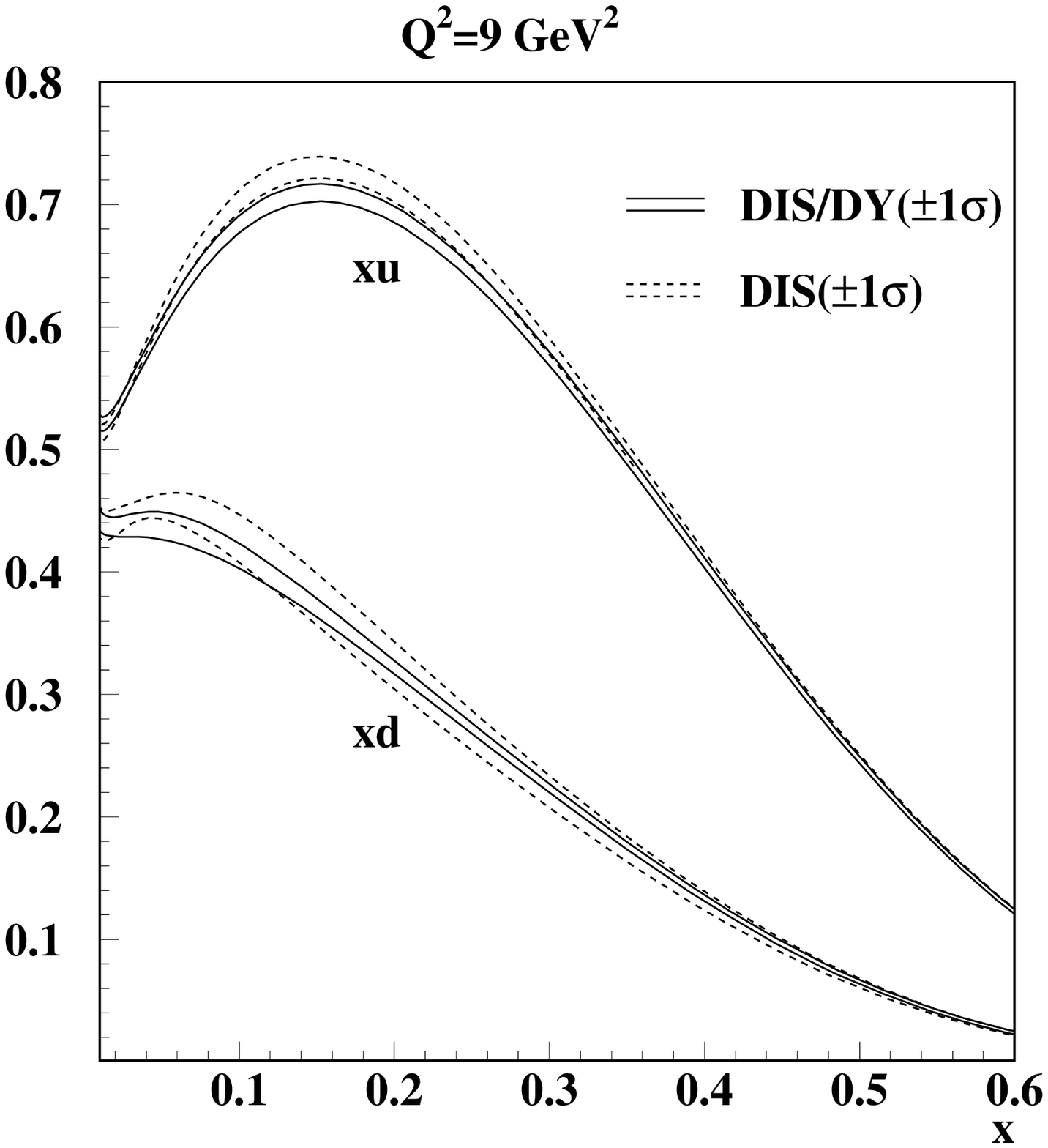,height=6.0cm,width=8cm,angle=0}
\psfig{figure=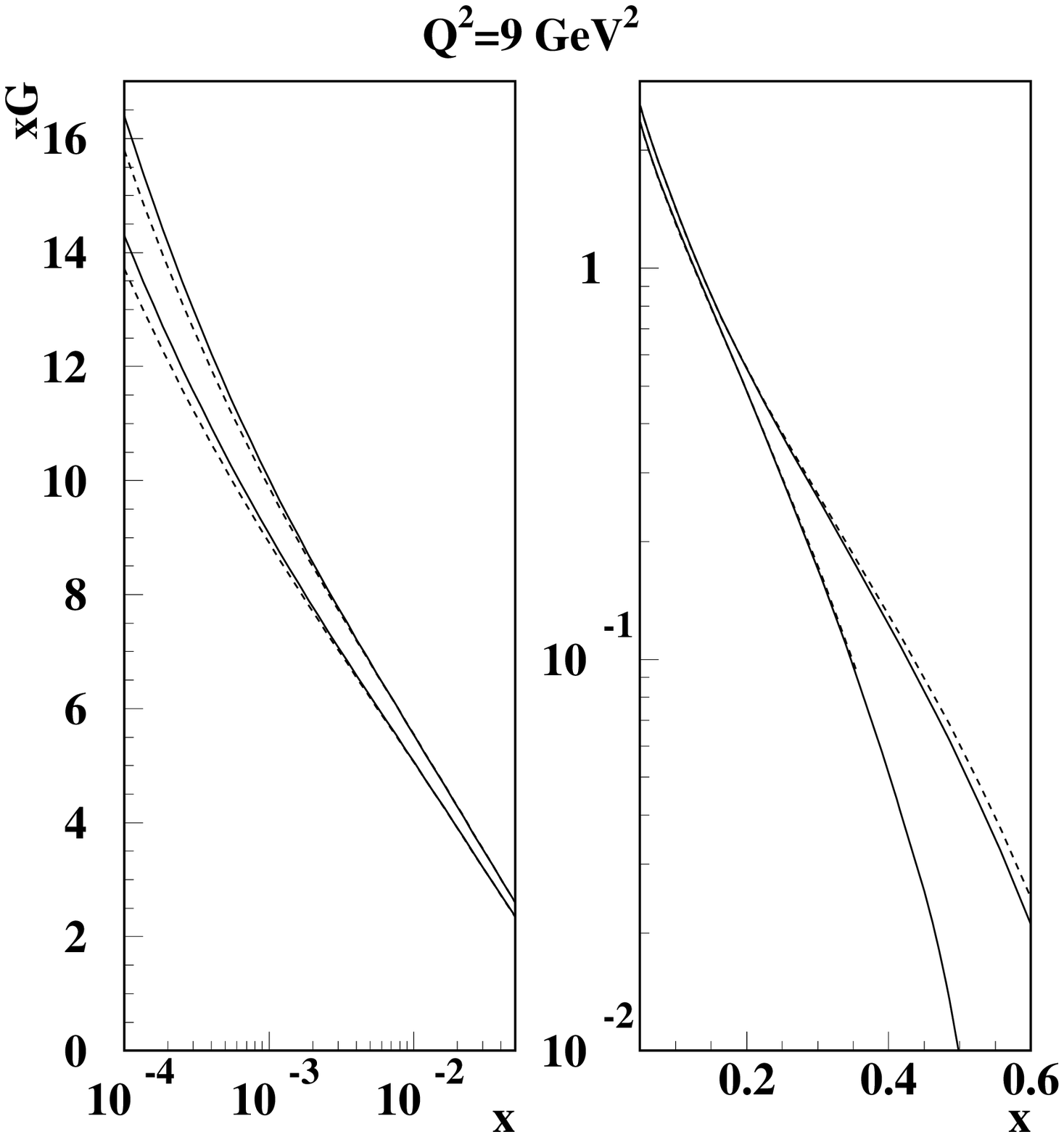,height=6.0cm,width=8cm,angle=0}
}
\caption{The same as in Fig.\ref{qs} for up (down) quarks and gluons.
}
\label{qvslx}
\end{figure}

Having established that the combined DIS/DY fit leads to an acceptable 
description of the data, we discuss the major differences between  the DIS/DY and 
DIS PDFs. For this comparison, the DIS PDFs were re-calculated 
using  the parameterizations shown in 
Eqs.~(\ref{eqn:pdf1}-\ref{eqn:pdf5}). Hence, the 
comparison presented below illustrates the differences in PDFs 
caused by the inclusion of the fixed target 
Drell-Yan data into the fit.

As we discussed in Section~\ref{sec.DIS-DY}, we expect  sea quark 
distributions at large values of $x$ to be  mostly affected by the DY data.
This is indeed what happens, as shown in Fig.~\ref{qs}. Both the 
symmetric and anti-symmetric combinations of $\bar u$ and $\bar d$ 
distributions are displayed. 
Dramatic improvements in the precision for large values of $x$ are observed 
once  the DY data are included in the fit.
For $x \lesssim 0.1$, the impact of the DY data on the isospin-symmetric 
combination $x(\bar{u} + \bar{d})$ is marginal, whereas the 
precision of the combination 
$x(\bar{d} - \bar{u})$ in the DIS/DY fit is higher for  $x>0.02$.  The central 
values of the sea quark distributions obtained in the DIS/DY and DIS 
fits agree within the errors, indicating 
consistency between the DIS and DY data. 
The largest discrepancies  
are at the level of one standard deviation; they  
occur at small $x$, where  the DIS and  DY data have comparable 
precision.

A better separation of  sea and valence quark distributions
in the DIS/DY fit leads to  an increased  precision of quark distributions, 
as shown in Fig.~\ref{qvslx}. The  effect is more pronounced for 
the $d$-quark content of the proton.
Both the $u$- and $d$- distributions obtained in the DIS/DY fit 
are smaller than similar distributions in  
the DIS fit at moderate values of $x$, but the difference is about $1\sigma$.
The gluon distribution is practically
unaffected by the DY data used in the fit, as seen in Fig.~\ref{qvslx}. 

\noindent
\begin{figure}[htbp]
\vspace{0.0cm}
\centerline{
\psfig{figure=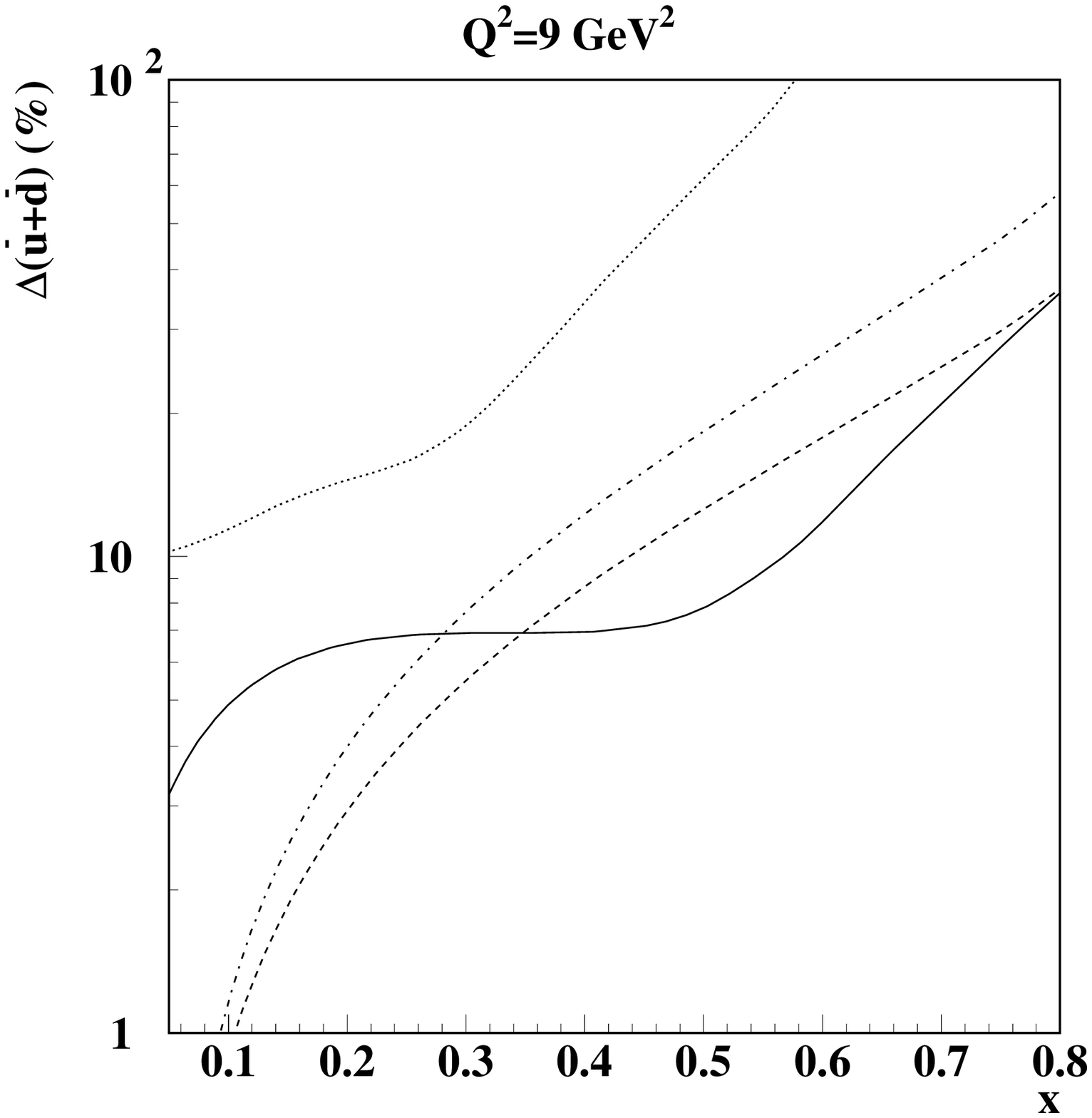,height=6cm,width=8cm,angle=0}
\psfig{figure=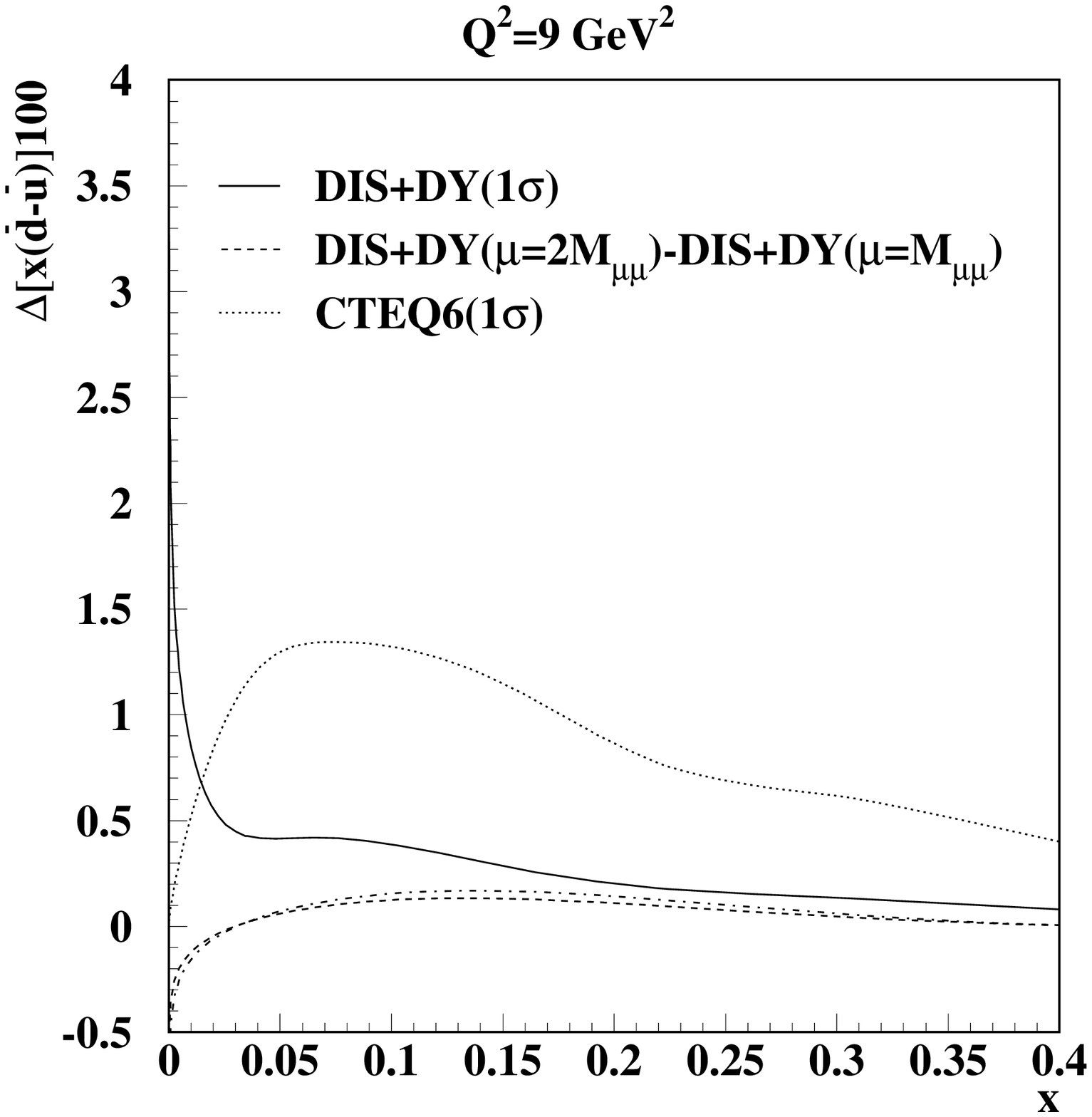,height=6cm,width=8cm,angle=0}
}
\caption{The 1$\sigma$ errors on the 
isospin symmetric and anti-symmetric sea quark distributions
due to  uncertainties in data. The results of the current   analysis (solid) 
are compared to that of  the CTEQ collaboration (dots) and to the 
uncertainties due to variations of the renormalization ans factorization 
scales (dashes). The latter quantity with the DY cross-section calculated 
through NLO in perturbative QCD is also given for comparison (dot-dashes).
}
\label{err}
\end{figure}

The theoretical errors of the DIS/DY PDFs  due to variations
of the renormalization and factorization scales do not exceed the 
``experimental'' errors  obtained by propagating  statistical and systematic
uncertainties in the data. The theoretical  uncertainties have the largest 
impact on the isospin-symmetric combination of sea quark distributions, where 
the theoretical and experimental errors are comparable
for $x\gtrsim 0.3$; this is shown in Fig.~\ref{err}.  
The  DIS/DY fit constrains
non-strange sea quark distributions with a precision better than   
$\pm 30\%$ for $x \sim 0.7$. The NNLO QCD corrections to the DY process 
are crucial for achieving this precision. If the NLO QCD 
theoretical prediction  for the DY rapidity distribution is used 
in the fit, the theoretical uncertainty due to the renormalization
scale variation is a factor of two larger than in the 
NNLO fit; as shown in Fig.~\ref{err}, it exceeds experimental errors in the 
isospin-symmetric sea distribution at large values of 
$x$.

The similar error estimated  
in the CTEQ fit \cite{cteq} is an order 
of magnitude larger, as shown in Fig.~\ref{err}. One of the reasons for
this disagreement is that in the CTEQ analysis, 
the criterion $\Delta\chi^2=100$
is applied to  account for possible inconsistencies in the 
data. In our case, good data consistency  is  a pre-requisite 
for assembling the data sample. Hence, we  apply the 
standard criterion $\Delta\chi^2=1$ that allows us  
to use the full power of the statistical analysis in our PDF determination.

\subsection{Phenomenological implications}

In this Section we briefly discuss some phenomenological implications of the 
above analysis.  A broad measure of the consistency of PDFs 
with other observables is provided by the value of the strong 
coupling constant $\alpha_s(M_{\rm Z})$.  The 
strong coupling constant obtained in the DIS/DY fit, 
$\alpha_s(M_{\rm Z}) = 0.1128(15)$, agrees with the value obtained in the DIS fit of Ref.~\cite{alekhin} 
within errors.
It is interesting that PDF fits generally prefer {\it smaller} values 
of the strong coupling constant than the current world average value 
$\alpha_s(M_{\rm Z}) = 0.1176(20)$ \cite{PDG}, 
and that the inclusion of NNLO corrections into the 
fits makes the disagreement {\it larger}
(see also the recent results of Ref.~\cite{Blumlein:2004ip}). 

Another interesting observable to discuss is the  Pascos-Wolfenstein 
ratio.  Recently,  the NuTeV collaboration measured 
the Weinberg angle in 
neutrino-nucleon scattering~\cite{Zeller:2001hh} and observed 
an anomaly. The significance of this anomaly is still an open issue 
since its interpretation depends on subtle details  of the quark structure 
of the nucleon and on the correct application of  
QCD and electroweak radiative corrections
\cite{nutevan}.  While discussing these 
issues is beyond the scope of this paper,  we would like to illustrate briefly the importance of improving PDFs 
at large values of $x$ for the NuTeV analysis.

For the sake of illustration, we consider
the Pascos-Wolfenstein ratio 
\begin{equation}
R^-=\frac{\sigma^\nu_{\mathrm{NC}}-\sigma^{\bar\nu}_{\mathrm{NC}}}
{\sigma^\nu_{\mathrm{CC}}-\sigma^{\bar\nu}_{\mathrm{CC}}}
\approx \frac12 -\sin^2 \theta_W.
\label{eq.pw}
\end{equation}
Although the NuTeV collaboration does not measure this ratio directly, we 
assume that $R^-$ is extracted from the data and is used to determine
the Weinberg angle.  The simple 
relation between $R^-$ and $\sin^2 \theta_W$ in Eq.~(\ref{eq.pw}) is only valid for an 
isoscalar target. Since the iron target used 
by NuTeV is not isoscalar, there is a correction to 
the Pascos-Wolfenstein 
ratio \cite{Kulagin:2003wz}
\begin{equation}
\delta R^{-}
\approx \frac{2Z-A}{A}\left(\frac{x_{1}^-}{x_{0}^-}\right)\left
(1-\frac{7}{3} \sin^2 \theta_W\right),
\end{equation}
where $A$ and $Z$ are the target atomic weight and charge and 
\begin{equation}
x_{0,1}^- = \int \limits_{0}^{1} 
\mathrm{d} x\: x(u_{\rm val}\pm d_{\rm val}).
\end{equation}
For iron , $\delta R^{-}$ is a factor of ten larger than the 
NuTeV experimental error; hence  the 
ratio ${x_{1}^-}/{x_{0}^-}$ must be known to better than 
10\%.  For the DIS/DY PDFs obtained in this paper,  the 
value of $x_{1}^-/x_{0}^-$ at $Q^2=20~{\rm GeV}^2$
is $0.4459\pm0.0094$. The DIS/DY PDFs therefore suppress the 
errors in the determination of $\sin^2 \theta_W$ due 
to the non-isoscalarity of NuTeV target to an acceptable 
value. We stress that inclusion of the DY data into the fit is crucial 
for achieving this accuracy. For example, in the NNLO DIS fit of Ref.\cite{alekhin} 
the value $x_1^{-}/x_0^{-} = 0.4324 \pm 0.0281$ at $Q^2 = 20~{\rm GeV}^2$
was obtained. In the global NLO fits by the CTEQ and MRST collaborations, these 
values are $0.4197 \pm 0.0307$ and $0.4317 \pm 0.0204$, respectively.

\begin{figure}[htbp]
\vspace{0.0cm}
\centerline{
\psfig{figure=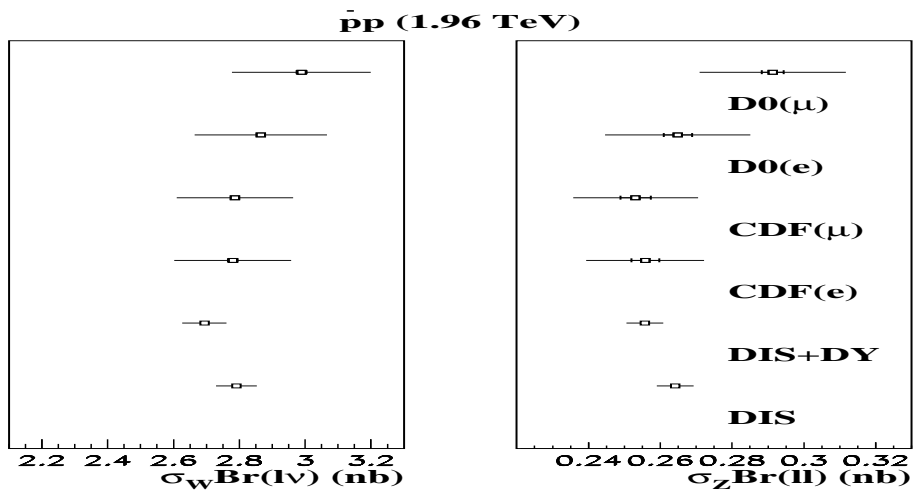,height=7.0cm,width=15cm,angle=0}
}
\caption{The preliminary Run II data for the $W$ and $Z$ production rates
measured at the Tevatron. The NNLO theoretical predictions  
predictions are obtained   with the DIS/DY PDFs and 
the DIS PDFs of Ref.~\cite{alekhin}.} 
\label{fnal}
\end{figure}

The production  of $Z$ and $W$ bosons 
at  hadron colliders can be used to measure 
partonic luminosities  \cite{Dittmar:1997md}.
In Fig.~\ref{fnal}, the NNLO QCD
predictions for these rates calculated using the   
DIS/DY PDFs  and DIS PDFs of Ref.~\cite{alekhin} 
and the coefficient functions 
of Ref.~\cite{Hamberg:1990np} are compared to  recent 
Tevatron results \cite{Bellavance:2005rg}. 
The errors in  the theoretical predictions arise from 
experimental uncertainties in the data 
used in the PDF fit; additional uncertainties 
come from varying the normalization factor $A_{\rm s}$ in 
Eq.~(\ref{eqn:pdf3}) by 40\% and from varying the charm quark mass by 
20\%.
Given the experimental errors on the$Z$ and $W$ production cross-sections,
the theoretical predictions agree with the measured rates.  
The theory results obtained with the DIS/DY and DIS PDFs agree within 
one standard deviation, demonstrating good stability of the 
fits with respect to the selection of data.  

\noindent

\section{Conclusions}

In  this paper we extend the NNLO QCD analysis of 
proton PDFs performed in Ref.~\cite{alekhin} 
by including fixed target Drell-Yan data into the fit. The possibility 
to do so without compromising the precision is due to the computation of the 
dilepton rapidity distribution in the DY process 
through NNLO in QCD \cite{dyrap,zrap}. When assembling the data sample, 
we pay particular attention to the consistency of the DIS and DY data.  We 
find that the DY data does not agree with the DIS data for large 
dilepton rapidities;  the disagreement actually becomes worse when 
the NNLO QCD corrections to the DY cross sections are taken into account.
 For this reason, we only include 
the E-605 data  and the E-866 data on the 
ratio of proton and deuteron cross-sections in the combined DIS/DY fit.

We find that the DY data improves the precision of sea quark PDFs 
at large values of  $x$, $x \gtrsim 0.1$. The overall quality 
of the DIS/DY fit is good, with $\chi^2/{\rm NDP} = 1.13$.  The differences 
between the DIS/DY PDFs obtained in this paper and the DIS PDFs derived in 
Ref.~\cite{alekhin} do not exceed one standard deviation, demonstrating 
good consistency of the data. 

{\bf Acknowledgments} We are grateful to S.~Kulagin and R.~Petti for 
useful discussions. S.A. is partially supported 
by the RFFR grant 06-02-16659 and 
the Russian Ministry of Science and Education 
 grant Nsh 5911.2006.2. K.M. is supported 
in part by the DOE under grant number DE-FG03-94ER-40833, Outstanding Junior 
Investigator Award and by the Alfred P. Sloan Foundation. F.P. is supported 
in part by the University of Wisconsin Research Committee with funds 
provided by the Wisconsin Alumni Research Foundation.


\end{document}